\documentclass[twocolumn,twocolappendix,usenatbib]{aastex631}

\def\eg{{\it e.g.}}

\def\etc{{\it etc.}}
\def\ie{{\it i.e.}}

\def\Msun{M$_\odot$}

\def\pmb#1{\setbox0=\hbox{$#1$}%
  \kern-0.25em\copy0\kern-\wd0
  \kern.05em\copy0\kern-\wd0
  \kern-0.025em\raise.0433em\box0}
\def\spmb#1{\setbox1=\hbox{${\scriptstyle #1}$}%
  \kern-0.25em\copy1\kern-\wd1
  \kern.05em\copy1\kern-\wd1
  \kern-0.025em\raise.0433em\box1}

\def\half{{\leavevmode\kern.1em\raise.5ex\hbox{\the\scriptfont0 1}\kern-.1em
    /\kern-.1em\lower.25ex\hbox{\the\scriptfont0 2}}}
\long\def\Ignore#1{\relax}

\def\spose#1{\hbox to 0pt{#1\hss}} % from Scott Tremaine
\def\gtlt{\mathrel{\spose{\lower.5ex\hbox{$\mathchar"13E$}}
     \raise.5ex\hbox{$\mathchar"13C$}}}
\usepackage{color}
\definecolor{red}{rgb}{0.7,0.1,0.1}
\definecolor{blue}{rgb}{0.2,0.2,0.8}
\definecolor{green}{rgb}{0.1,0.6,0.1}

\graphicspath{{./}{figures/}}
\begin{document}

\title{After 54 years of bar instability studies: a fresh surprise}
\shorttitle{Bar stability}

\author{J. A. Sellwood}
\affiliation{Steward Observatory, University of Arizona,
933 Cherry Avenue,
Tucson, AZ 85722, USA}
\email{sellwood@arizona.edu}

\author{Victor P. Debattista}
\affiliation{Jeremiah Horrocks Institute, University of Central Lancashire,
Preston, PR1 2HE, UK}
\email{vpdebattista@gmail.com}

\author{R. G. Carlberg}
\affiliation{Department of Astronomy and Astrophysics, University of Toronto,
Toronto, ON M5S 3H4, Canada}
\email{raymond.carlberg@utoronto.ca}

\shortauthors{Sellwood, Debattista and Carlberg}

\centerline{To appear in ApJ}

\begin{abstract}
The well-known bar instability of rotationally-supported disk galaxy
models has been studied extensively since its first discovery over
mvhalf a century ago.  We were therefore very surprised to find cases
of disks embedded in rigid halos, which on the basis of widely-cited
criteria should be unstable, that appeared to be robustly stable.
Here we show that the unstable bar mode in such simulations was being
suppressed by changes to the disk caused by other instabilities having
higher angular symmetry that were the first to saturate.  Although
this may seem like a promising solution to the long-standing puzzle
presented by the apparent stability of real disk galaxies, we also
show that instability is restored in the same models when the rigid
halo is replaced by a live population of particles, where the usual
stability conditions apply. Our study has been confined to a narrow
range of models, and we cannot therefore exclude the possibility that
mode interference may be able to prevent bar formation in other models
having live halos.
\end{abstract}


\section{Introduction} \label{sec.intro}
\citet{Hohl71} presented perhaps the first careful simulations that
revealed the tendency for disk galaxy models to undergo a global
instability that rearranged a rotationally-supported disk of stars
into a strongly barred configuration.  Soon thereafter \citet{OP73}
suggested that the survival of nearly axisymmetric disk galaxies may
require them to be embedded in a halo of dark matter, provoking many
follow up studies in both theory (\eg\ \citealt{Zang76},
\citealt{Kaln78}, \citealt{Toom81}) and numerical work
(\eg\ \citealt{CS81}, \citealt{CST95}, \citealt{AM02}, \citealt{DCMM})
that has continued to this day \citep[\eg][]{SC23}.  Despite all these
studies, we still lack a widely-accepted explanation for the apparent
stability of isolated disk galaxies.

It is to be hoped that we may one day identify the explanation for
which galaxies do or do not host bars from galaxy formation
simulations.  These massive experiments \citep[see][for a
  review]{FB25}, that mimic many physical processes, are increasingly
able to synthesize somewhat realistic model disk galaxies, especially
those of Milky Way mass, but the fraction of these models that support
bars, and their properties, vary widely between the different codes
that are employed \citep[\eg][]{Algo17, Zhao20, Rosa22, Redd22,
  Ande24, Frag25, Ansa25, Lu25}.  Though studies of isolated galaxies
have provided helpful guidance on the evolution of bars
\citep[\eg][]{Ansa25}, the very complexity of the physical processes
in the cosmological context have thus far left the experimenters
\citep[\eg][]{Zhou20, Ansa25} conceding that they are unable to
identify the mechanism or properties -- star formation and feed back,
disk/halo mass fractions, tidal encounters, mergers, \etc\ -- that
caused a particular galaxy model to host a bar.  Both the algorithms
and computer power continue to improve, which it is hoped will
eventually enable this question to be answered.  In the meantime we
pursue a parallel investigation using idealized models in which we
have some hope of developing deeper insight into this complicated
question of disk dynamics.

\citet{ELN} undertook a systematic study of a family of galaxy models
having an exponential disk of mass $M_d$ and radial scale length $R_d$
embedded in various rigid halos, and reached the widely-cited
conclusion that the disk could avoid forming a bar only if the maximum
rotational velocity of the disk material, $V_m \ga 1.1
(GM_d/R_d)^{1/2}$.  Since the maximum circular speed due to the disk
alone is $\sim 0.62(GM_d/R_d)^{1/2}$ \citep{Free70}, they argued that
stable galaxies must be embedded in massive halos to make up the
required circular speed.

The surprise we report here (\S2) is counter examples that violate
their stability criterion for a reason that, to our knowledge, has not
previously been identified: changes to the equilibrium disk caused by
saturation of faster growing spiral modes that disturbed the incipient
bar-forming mode halting its linear growth at an early stage.  The
apparently stable models were reruns of some of those simulated by
\citet[hereafter SC23]{SC23}, but which included force terms from
multiple sectoral harmonics.  As a result of our discovery, we were
greatly concerned that the principal finding from SC23, that all their
models were bar unstable, suddenly appeared to be incorrect because it
had been based on simulations that were restricted to $m=2$ only
disturbance forces.

Suppression of bars by prior saturation of competing modes may seem
like a promising solution to the bar-instability problem highlighted
by \citet{OP73} and by \citet{ELN}.  However, we further report here
that the apparent stability of our particular disk galaxy models is a
numerical artifact resulting from employing rigid halos.  Replacing
the rigid halo by a similar one composed of mobile particles allowed
the disk to form a strong bar, as expected.  While this finding
suggests that almost all the models studied by SC23 were indeed
unstable, as they had claimed, it does not rule out the possibility
that disks in other live halo models could be stabilized by similar
non-linear changes by faster growing modes.

\newpage
%%%%%%%%%%%%%%%%%%%% Section 2 %%%%%%%%%%%%%%%%%%
\section{Models and methods}
\subsection{Rigid halo models}
The two galaxy models we employ in this section were selected from the
set used by SC23.  The exponential disk has
the surface density profile
\begin{equation}
  \Sigma(R) = \Sigma_0 e^{-R/R_d} \quad\hbox{with}\quad \Sigma_0 = {M_d
    \over 2\pi R_d^2},
\label{eq.expdisk}
\end{equation}
where $R_d$ is the disk scale length and $M_d$ is the nominal mass of
the infinite disk.  We limit its radial extent using a cubic function
to taper the surface density from $\Sigma(5R_d)$ to zero at $R=6R_d$.

The rotation curve is that of a cored isothermal sphere
\begin{equation}
  V_c(R) = V_0 \left[ { R^2 \over R^2 + r_c^2} \right]^{1/2},
\label{eq.Vcirc}
\end{equation}
with $r_c$ being the core radius.  The implied halo density is
whatever is required, when combined with the disk attraction, to
achieve this rotation curve in the disk plane \citep{FE80}.  We relate
the rotation curve to the disk properties by setting $V_0 = 0.9
(GM_d/R_d)^{1/2}$, and choose $r_c = R_d/2$ for model A, as the
baseline model of SC23, and $r_c = R_d$ for model B.  Though quite
heavy, the disk has less than the required mass to account for the
central attraction at all radii.  SC23 reported that these two, and
nearly all other models in their study, possessed vigorous, global bar
instabilities.

As usual, we adopt units that $G = M_d = R_d = 1$, so a dynamical time
is $(R_d^3/GM_d)^{1/2}$, \etc \ For those who prefer physical units,
one possible scaling is to choose $R_d=2.5\;$kpc and the dynamical
time to be 10~Myr, which implies $M_d = 3.47 \times 10^{10}\;$\Msun,
and a velocity unit $(GM_d/R_d)^{1/2} = 244.5\;$km~s$^{-1}$.

SC23 employed the method proposed by \citet{Shu69}, with numerical
details given in \citet{Sell14}, to create an equilibrium distribution
function (DF) for the disk particles, which for both models had
$Q=1.2$ at all radii. The sense of net rotation in all models
presented in SC23 was positive at all radii, but those authors avoided
a discontinuity in the DF by flipping the sign of $L_z$ for a small
fraction of low $L_z$ particles.

\begin{table}
\caption{Default numerical parameters for our 2D simulations, the last
  four of which are independent of the grid type.}
\label{tab.gpars}
\begin{tabular}{@{}lll}
  & Polar grid & Cartesian grid\\
\hline
Grid points & 171 $\times$ 256 & 1024 $\times$ 1024 \\
Scaling to grid units & $R_d= 10$ & $R_d = 80$ \\
Active sectoral harmonics & $1 \leq m \leq 8$ & unrestricted \\
\hline
Plummer softening length & $\epsilon = R_d/20$ \\
Number of particles & $6 \times 10^6$ \\
Largest time-step & $0.2R_0/V_0$ \\
Radial time step zones & 5 \\
\end{tabular}
\end{table}

\subsection{2D simulation codes} \label{sec.code}
We select particles from the adopted DF using the procedure described
in \citet{Sell24}, place them in a plane at random azimuths
(\ie\ a noisy start, see \S\ref{sec.starts} below) and compute
the mutual attractions of the particles using either a 2D polar, or a
2D Cartesian, mesh.  This code is described in detail in
\citet{Sell14}; in summary, the particles move subject to forces from
other particles that are interpolated from the grid.  We adopt the
parameters listed in Table~\ref{tab.gpars}.

In the polar grid simulations we report in this section, the central
attraction is that of a rigid halo needed to ensure centrifugal
balance $a_R=-V_c(R)^2/R$, and we neglect the axisymmetric part of the
attraction from the mobile particles. We also generally suppress
sectoral harmonics $m>8$ from the force determination, which would add
only noise, and describe forces from the active components $1 \leq m
\leq 8$ as ``unrestricted'', but we also report results from some
simulations is which non-axisymmetric forces were restricted to $m=2$
only. We include all force terms from the Cartesian grid, and
supplement the central attracton to maintain the same rotation curve.
We employ block time steps that are decreased by factors of 2 in each
radial zone.

As usual, we measure non-axisymmetric distortions of the distribution
of the $N$ disk particles using an expansion in logarithmic spirals:
\begin{equation}
A(m,\gamma,t) = {1 \over N}\sum_{j=1}^N \, \exp[im(\phi_j + \tan\gamma \ln R_j)],
\label{eq.logspi}
\end{equation}
where $(R_j,\phi_j)$ are the polar coordinates of the $j$th particle
at time $t$, $m$ is the sectoral harmonic, and $\gamma$ is the
(radially constant) angle of the spiral component to the radius
vector, which is the complement to the spiral pitch angle.

\begin{figure}
\begin{center}
  \includegraphics[width=\hsize,angle=0]{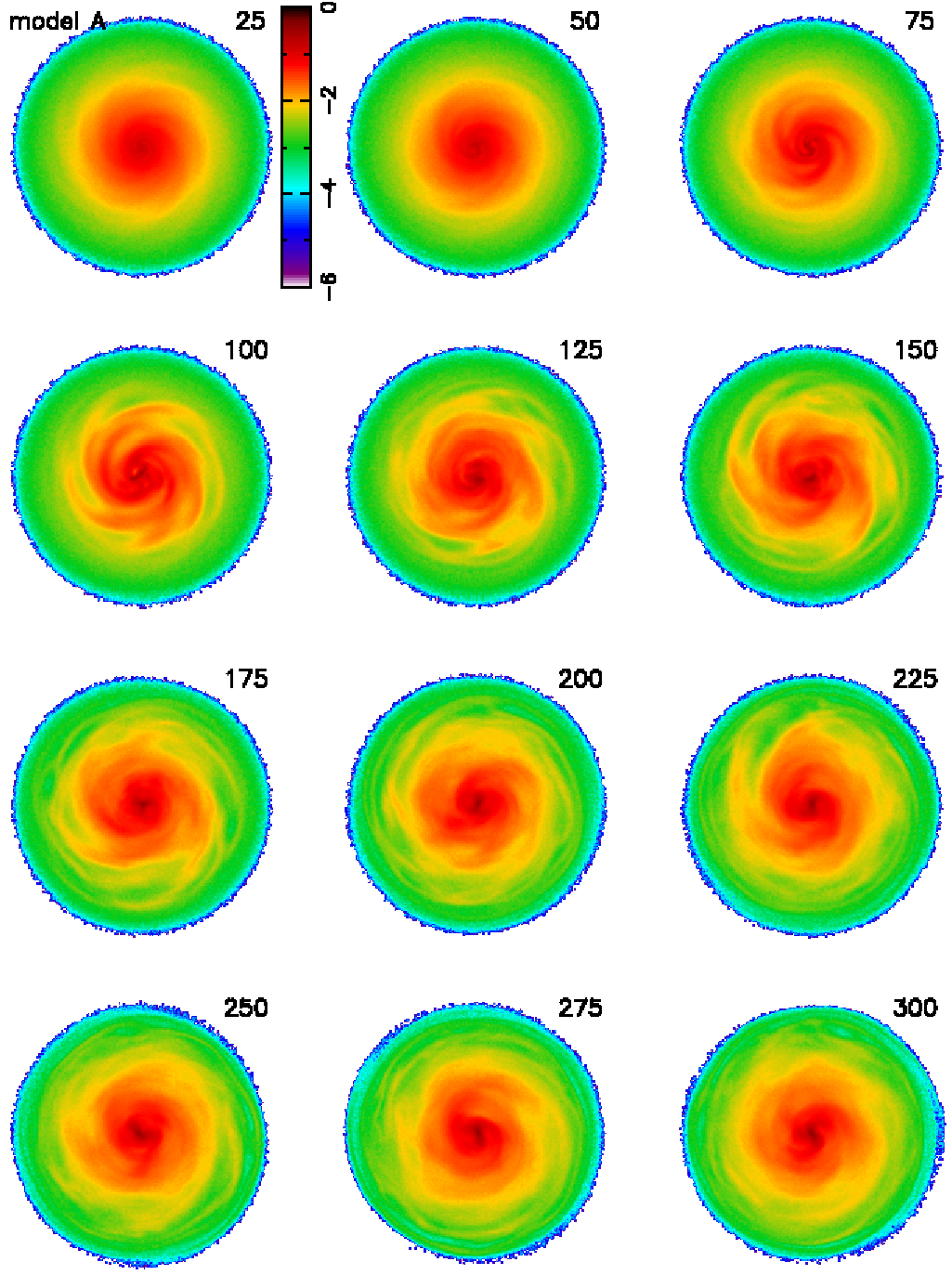}
% jasmine:/data/sellwood/5700/5772/proj.s 5772
\end{center}
\caption{The evolution of a noisy start realization of the baseline
  model from SC23 with force terms $1 \leq m \leq 8$ all active.
  The color scale indicates the logarithm of the disk surface density.
  Notice that no strong bar forms, though a short, weak bar is visible
  from time to time.}
\label{fig.run5772}
\end{figure}

\subsection{Noisy and quiet starts}
\label{sec.starts}
Random selection of the initial particle coordinates from the
desired distributions creates an initial model in which density
fluctuations are caused by undiminished shot noise arising from the
finite number of particles. Evolving such a ``noisy start'' model
with unrestricted forces allows all possible modes, both neutral and
unstable, to develop simultaneously.

\citet{Sell83} described how to create a ``quiet start,'' in which
particles are placed almost perfectly symmetrically on rings and given
identical orbital and radial velocities.  If non-axisymmetric
disturbance forces are also restricted to a single sectoral harmonic
the forces experienced by the particles are those from a smooth,
massive ring that distorts as expected from growing large-scale
disturbances while the initial regular arrangement inhibits
small-scale disturbances.  These tricks reduce the level of shot noise
by many orders of magnitude and enable identification of linear
instabilities that emerge and grow through several $e$-folds before
saturating. See \citet{Sell24} for more details.

\subsection{A noisy start simulation} \label{sec.noisy}
Fig~\ref{fig.run5772} shows the evolution of a noisy start version,
with sectoral harmonics $1\leq m\leq8$ all active, of model A, which
is the baseline model described in SC23.  The disk manifested multi-arm
spiral patterns but did not form a bar.

This figure is to be compared with Fig~3 of SC23 which revealed that
the $m=2$ only, quiet start version of the same model that started
from the same file of particles, formed a large, strong bar.  Note
that the $e$-folding time of the dominant $m=2$ mode reported in SC23
is $\sim 32$ dynamical times, or $\la 1/9$ of the interval illustrated
in Fig~\ref{fig.run5772}, which is ample for the bar to have emerged
from the noisy start if the same mode had saturated in this new
simulation.

We have verified that a rerun of the same model using a Cartesian
grid with unrestricted forces did not form a bar either and its
evolution closely resembled that illustrated in Fig.~\ref{fig.run5772}.

\begin{figure}
\begin{center}
  \includegraphics[width=.8\hsize,angle=0]{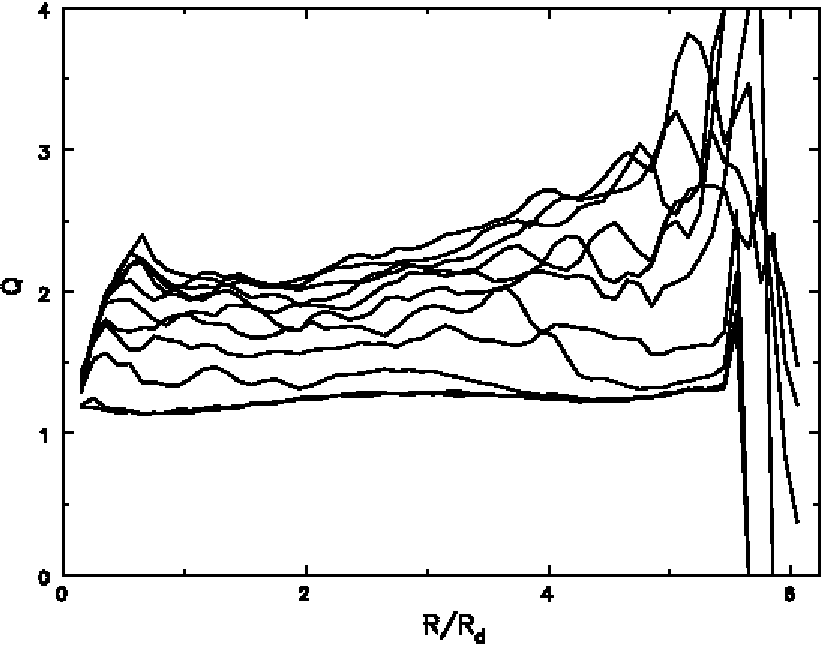}
% jasmine:/data/sellwood/5700/5772/QRt.s 5772
\end{center}
\caption{The radial variation of $Q$ at intervals of 50 dynamical
  times in the simulation illustrated in Fig.~\ref{fig.run5772}.  The
  lines are not labeled because increases in $Q$ are mostly
  monotonic.}
\label{fig.QRt5772}
\end{figure}

It has long been known \citep[\eg][]{SC84} that spiral activity heats
the disk and causes the spirals to fade over time.
Fig.~\ref{fig.QRt5772} presents the time evolution of the radial $Q$
profile in model A revealing a rise to $Q \ga 2.3$ over the range $1
\leq R \leq 4$, and still higher at larger radii.  Heating of the very
inner disk was largely suppressed by the high density of the inner
halo.

\subsection{Comparison with ELN}
\label{sect.ELN}
The pioneering simulations by \citet{ELN}, hereafter ELN, were similar
in almost all respects to the noisy start model illustrated here as
Fig.~\ref{fig.run5772}, except that the limited computational power
available at the time forced them to employ merely $2 \times 10^4$
particles and a 2D grid that had $128^2$ cells.  From these crude, by
today's standards, simulations, they concluded that bars always formed
provided $V_0 \la 1.1$ (eq.~\ref{eq.Vcirc}).  We note that our
simulation illustrated in Fig.~\ref{fig.run5772} had $V_0 = 0.9$.  A
noisy start simulation of our model B, which also had $V_0 = 0.9$ but
the larger core radius $r_c = R_d$, did not form a strong bar either,
and this model had exactly the same properties as Model 11 in the
paper by ELN.  Those authors concluded from the slow decline of their
$\delta_2$ parameter (bottom, middle panel of their Fig.~2) that their
model 11 was bar unstable.

\begin{figure}
\begin{center}
  \includegraphics[width=\hsize,angle=0]{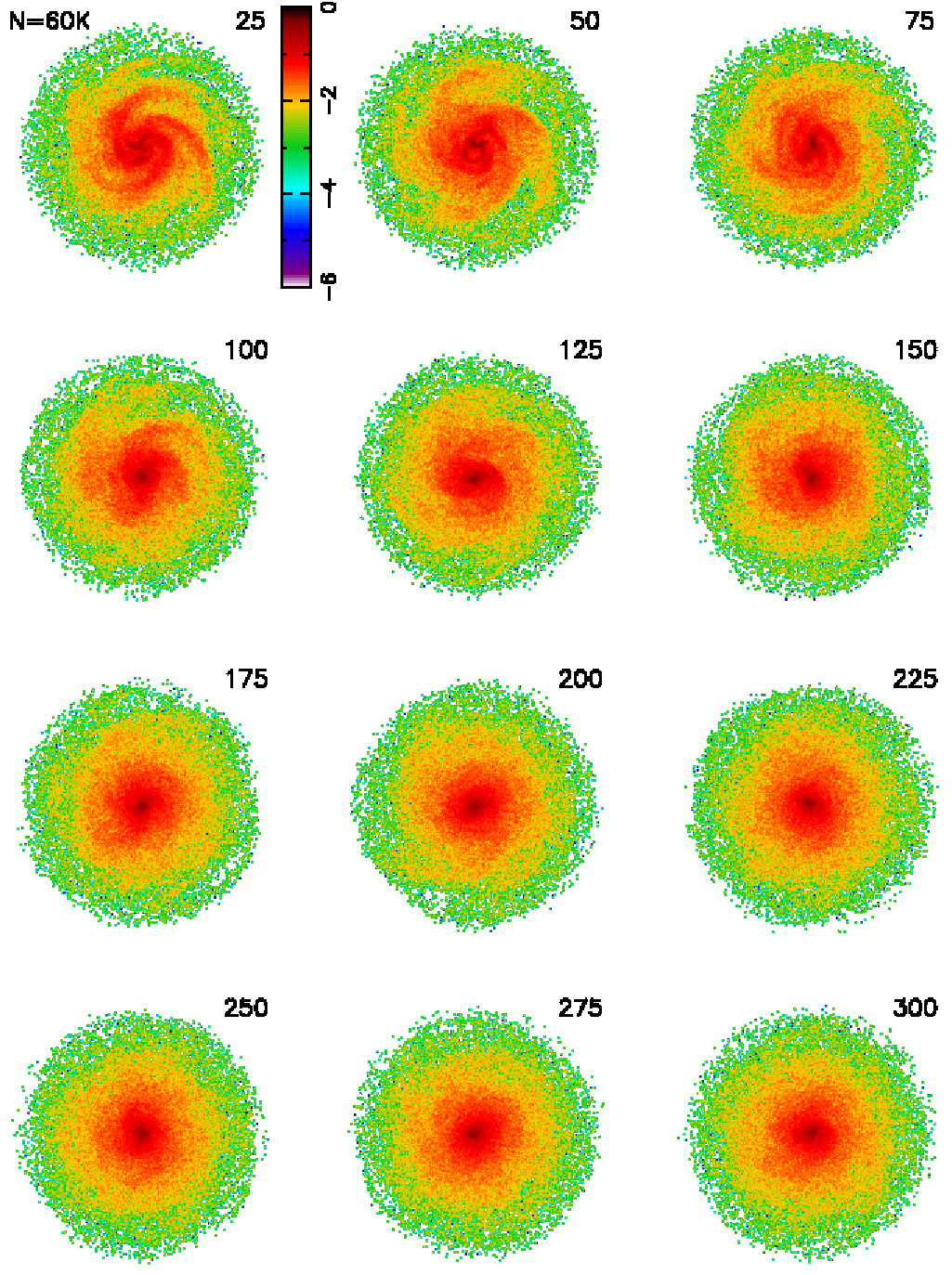}
% jasmine:/data/sellwood/5900/5915/proj.s 5916
\end{center}
\caption{The evolution of one comparison simulation of model A having
  $N=60$K particles.  The initial disk extends to $R=5R_d$ and times
  are given in dynamical times.}
\label{fig.proj5916}
\end{figure}

\begin{figure}
\begin{center}
  \includegraphics[width=.8\hsize,angle=0]{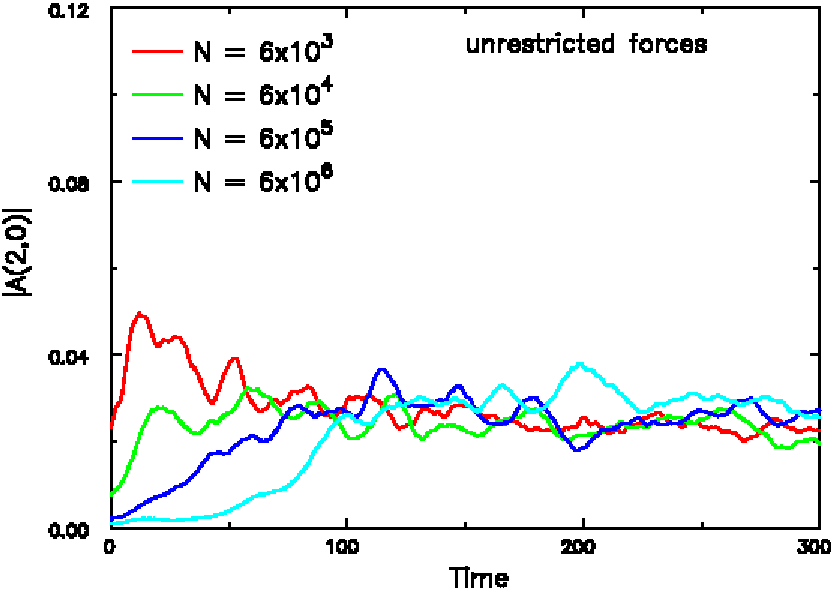}
  \includegraphics[width=.8\hsize,angle=0]{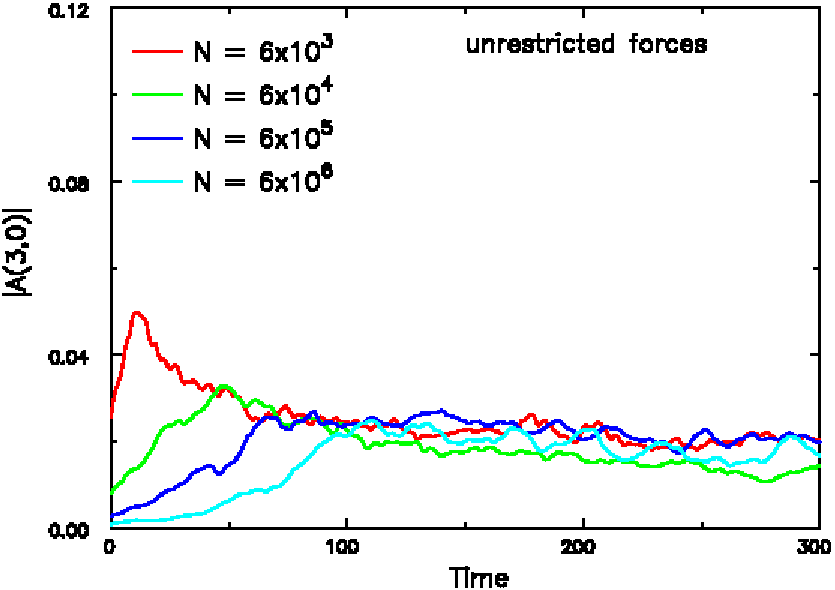}
  \includegraphics[width=.8\hsize,angle=0]{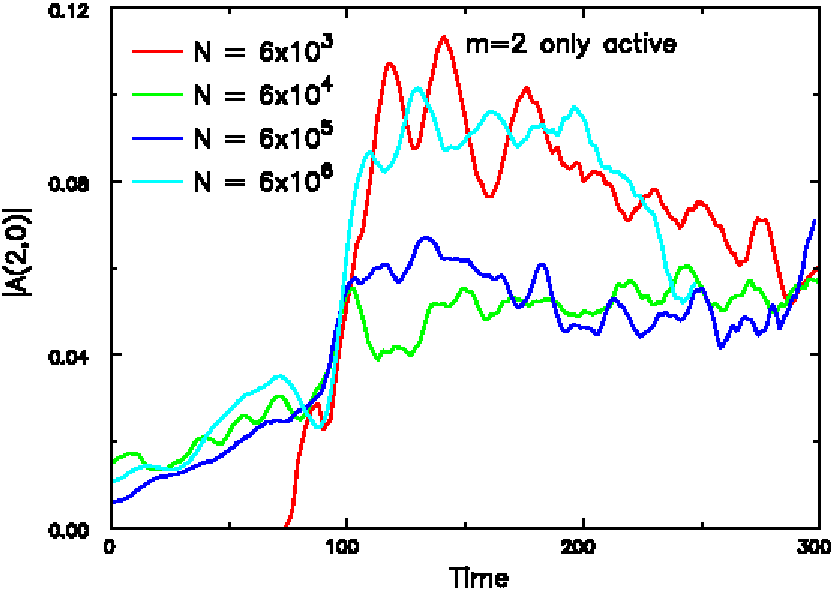}
% jasmine:/data/sellwood/5900/5934/means(1-3).s
\end{center}
\caption{Top: the mean amplitudes of the bar-like $A(2,0,t)$
  (eq. \ref{eq.logspi}) in 10 noisy start realizations of model A, for
  4 different values in $N$ with unrestricted forces.  Middle: the
  same as for the top panel, but for $m=3$.  Bottom: the evolution of
  the bar amplitude in identical noisy start simulations when
  non-axisymmetric forces are restricted to $m=2$.  Results from
  the individual simulations were shifted in time in the bottom panel
  only so that $A=0.06$ at $t=100$ for each before averaging.}
\label{fig.means}
\end{figure}

Neither of our noisy start simulations of models A and B, with
$V_0=0.9$, formed bars, which is in violation of the widely-cited
stability criterion proposed by ELN.  In order to investigate the
cause of this discrepancy, we have run a further 79 separate
simulations of model A, in which we employed four differing numbers of
particles, each with ten different random seeds used to select the
initial particle coordinates, and finally we computed all 40 models
twice, once with forces from sectoral harmonics $1 \leq m \leq 8$ all
active, and again with $m=2$ forces only.  Fig.~\ref{fig.proj5916}
illustrates the evolution of one case: the strong $m=3$ spiral pattern
that is already visible at $t=25$ is typical of all our low-$N$
simulations having unrestricted forces.

Each line in each panel of Fig.~\ref{fig.means} presents the time
evolution of the given logarithmic spiral amplitude, averaged over the
minor stochastic differences between ten realizations having different
random seeds, and the four lines in each panel span a factor of 1000
in the values of $N$.

The simulations in the top two panels employed unrestricted forces and
we report the amplitude of the bar-like $A(2,0)$ and the trefoil
$A(3,0)$ coefficients in the top and middle panels respectively.
Amplitudes are largest in the smallest $N$ simulations (red curves) at
early times, reflecting the higher level of seed noise, but there is
no evidence that the amplitudes at later times depend on the number of
particles employed.  The final bar amplitudes are $A(2,0) \sim 0.02$,
which is very weak; $A(2,0) \ga 0.12$ for strong bars (\eg,
\citealt{SA86}, SC23).  Notice also that the $m=3$ coefficients
decrease somewhat over time, perhaps exceeding those of $m=2$ at
first, but ending slightly smaller.  The behavior of the red line in
the top panel ($m=2$) resembles the evolution of the similar
$\delta_2$ parameter presented by ELN in their Fig.~2 for their
low-$N$ simulation, despite the different scaling.

The bottom panel of Fig.~\ref{fig.means} reports results from a
separate set of simulations starting from the same files of particles
as those in the top two panels, but in which we restricted
non-axisymmetric forces to the $m=2$ sectoral harmonic only. As
we suspected that a dominant bar-forming mode was present in these
runs (as found by SC23) we attempted to align the amplitude variation
to pass through $A=0.06$ at $t=100$ in each separate simulation before
averaging so that time offsets between the separate cases did not
obscure the trends.  This strategy was moderately successful, and it
is apparent that bars in these models have much greater amplitude than
in the top panel, a clear indication that non-linear interference from
modes having other symmetries inhibited the bar instability.

We followed up this hint by conducting other quiet start simulations
of model A with disturbance forces restricted to $m=3$ or $m=4$ to
search for other linear instabilities, finding one $m=4$ mode whose
growth rate was only slightly lower than that of the dominant $m=2$
mode, and two 3-fold symmetric modes having growth rates that exceeded
it.  All these, and probably other, modes would have started to grow
from the outset of the simulation illustrated in
Fig.~\ref{fig.run5772}.  Multiple spiral arms can be seen in this
figure as early as $t=100$, when the condition for independent linear
growth of each mode is clearly no longer
satisfied.\footnote{Namely that for as long as non-axisymmetric
  distortions and changes to particle velocities continue to be
  negligible.}  We therefore conclude that the bar instability can be
inhibited by non-linear changes caused by other modes, provided
they disturb the inner disk before the bar-forming instability
saturates.  We describe more fully what we mean by this statement in
\S\ref{sect.tech}.

Before this study, we were unaware of the rapidly growing additional
$m=3$ and $m=4$ modes that saturated before the bar mode, inhibiting
the large bar that developed in simulations only when disturbance
forces are restricted to $m=2$.  Our re-simulations of a few of the
models from ELN with $N=2 \times 10^4$ particles and $128^2$ Cartesian
grid, and also with the polar grid, possessed $m=3$ features that were
stronger than those of $m=2$ in the early evolution (see also
Fig.~\ref{fig.proj5916}), but the relative amplitudes of the two
sectoral harmonics was reversed after $t\sim 100$, consistent with the
persistence of $m=2$ reported by ELN, which evidence may have formed
the basis for their stability criterion.

\subsection{Technical discussion}
\label{sect.tech}
Linear instabilities \citep[see \eg][for a fuller discussion]{SM22}
grow exponentially for as long as the amplitude of every disturbance
is small enough that terms in the collisionless Boltzmann equation
(CBE) that are second order in the perturbation amplitude can be
neglected.  Once this assumption breaks down, the dominant mode is
said to saturate; the neglected terms in the CBE begin to cause finite
changes to the initial equilibrium model and its exponential growth
ceases.  The changes to the equilibrium model are largest at the
principal resonances of the mode \citep{LBK}, and alter both the
angular momenta of the affected particles and, at the Lindblad
resonances only, increase their random energy, which is the reason
that the non-linear evolution of spiral instabilities causes disks to
heat, as shown in Fig.~\ref{fig.QRt5772} \citep[see also][]{SC19,
  Roul25}.  Stars that lose angular momentum at corotation as a
vigorous bar forming mode saturates, on the other hand, become trapped
into a tumbling bar, while gainers at this resonance move to larger
orbits.

The initial amplitude of each mode is determined by the spectrum of
shot noise from the coordinates of each particle.  Thus changing the
random seed used for the initial positions of each particle will
change the initial amplitude of every mode, although each will again
start to grow at the rates of the separate linear modes.  If, as here,
there are several instabilities having comparable growth rates, the
mode that saturates first could differ between runs having different
random seeds, leading to macroscopic stochastic differences in the
later evolution \citep[see][for an in depth study]{SD09}.  It seemed
possible that the absence of a large bar in Fig.~\ref{fig.run5772}
could be a consequence of an unlucky random seed, but this idea was
ruled out by the results presented in the top panel of
Fig.~\ref{fig.means}.  Not one of the ten randomly seeded models at
each $N$ formed a strong bar.

In the case of the simulations reported by ELN, in which multiple
instabilities saturate in quick succession, some trapping into a bar
could perhaps start to occur at about the same time as the disk is
being heated by spiral modes.  When resonances of separate
large-amplitude modes overlap, the dynamics can become chaotic
\citep{DW18}, made more so by the changing amplitudes of
both disturbances in this case.  Lindblad resonance scattering by a
spiral mode may inhibit some of the trapping that would have occurred
if the bar mode had been isolated, leading to a much weaker bar as the
model settles after these events.  Indeed, ELN's measurements of
their parameter $\delta_2$ are surprisingly small compared with those
for bar instabilities uncontaminated by competing spiral
modes \citep[\eg][]{SA86}.

The linear growth rates of all modes should be the same regardless of
the particle number but, as noted above, the seed amplitudes of all
modes in ELN's simulations would have been much higher than those in
our $N=6$ million particle models.  Therefore the modes would have
saturated much sooner after the start, leaving little time for the
small differences in growth rates to matter.

The situation is cleaner when the times of saturation of the separate
modes are well spaced, as occurred in our model shown in
Fig.~\ref{fig.run5772}.  Some spiral modes grow more rapidly than does
the bar mode, and therefore the equilibrium disk in which the bar mode
is growing linearly can suddenly be altered by the non-linear heating
caused by a spiral instability. It would be difficult to predict the
outcome of this event, but the absence of even a vestige of a large
bar in our simulation indicates that the linear growth of the
bar-forming instability is terminated at small amplitude before any
trapping occurs.

The corotation radius of the bar instability in model A is $\sim 2.1$
while the outer Lindblad resonances (OLRs) of the faster growing $m=3$
modes mentioned above are at $R \sim 1.80$ and $\sim 1.93$.  Thus
while the exponentially growing bar instability was still at small
amplitude, these, and possibly other, spiral modes will have saturated
and heated the disk at their Lindblad resonances \citep{LBK}, thereby
changing the properties of the disk near corotation of the bar mode
during its linear growth.  This interference apparently killed off the
bar-forming mode because no bar developed in Fig.~\ref{fig.run5772}.

\begin{figure*}
\begin{center}
  \includegraphics[width=.32\hsize,angle=0]{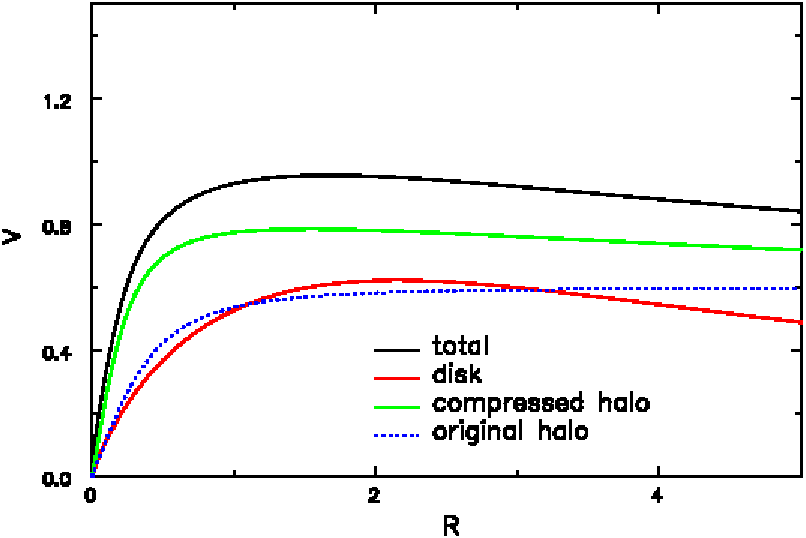}
  \includegraphics[width=.32\hsize,angle=0]{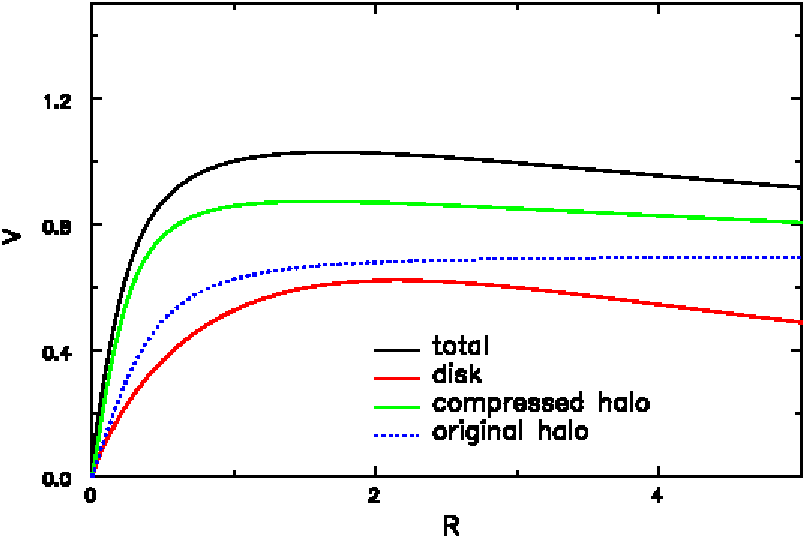}
  \includegraphics[width=.32\hsize,angle=0]{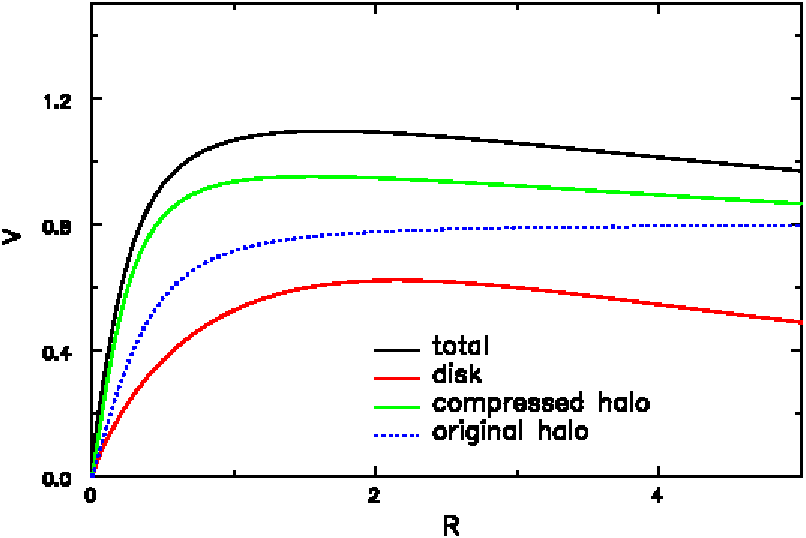}
% jasmine:/data/sellwood/5700/5772/picts.s 5772
\end{center}
\caption{Three example rotation curves (black curves) of disk +
  compressed halo.  The disk contribution, which is the same in all
  three is shown in red, and the compressed halo in green, with
  $V_0=0.6$ in the left panel, $V_0=0.7$ in the middle panel, and
  $V_0= 0.8$ in the right panel.  In all three panels the contribution
  of the halo before compression, for which $r_c = R_d/2$ is shown by
  the dotted curve.}
\label{fig.RCs}
\end{figure*}

%%%%%%%%%%%%%%%%%%%% Section 3 %%%%%%%%%%%%%%%%%%
\section{Live halo models}
It has long been known \citep{Atha02, SN13, BS16} that a live halo
provides a supporting response to the bar instability in a disk,
causing that mode to grow more rapidly than in the equivalent rigid
halo.  The open spiral of the bar mode couples strongly to those
bi-symmetric orbits in the halo that precess at similar rates to those
in the disk, especially any whose orbit planes are not far from the
disk plane \citep{Sell15}.  Thus the instability is that of the
combined halo+disk system, that a rigid halo could not capture.

In order to study the stability of a disk embedded in a live halo, it
is necessary first to create an equilibrium model.  However, it would
be extremely difficult to create live halo versions of the family of
simple models adopted by ELN, since the halo density in any one model
does not have a simple form.  Their models specify an attractively
simple functional form for the total rotation curve
(eq.~\ref{eq.Vcirc}), which results from the combined attraction of
the disk and halo.  The attraction of a thin exponential disk can be
expressed in terms of modified Bessel functions \citep{Free70}, but
the attraction of the halo must be that which would result from the
total radial attraction, $V_c^2(R)/R$, with the disk contribution
subtracted.  While the required halo attraction could be calculated
numerically, it clearly would not have a simple functional form.  Note
that not all possible models would be physically acceptable since a
heavy disk embedded in a halo having a large core radius $r_c$, could
require the halo density to be negative at some radii.

Possible methods that could be employed to construct an equilibrium DF
for the halo having an embedded disk are:

\begin{enumerate}
\item
\citet{Vasi19} proposed the {\sc agama} procedure to create an
equilibrium disk+halo model in any one case.  However, one must choose
a function of the actions, and the method converges to an equilibrium
model by iteration, making it hard to achieve exactly the desired
disk/halo mass ratio and rotation curve.

\item
A second option, adopted by \citet{HBWK}, is to use Eddington
inversion to find an equilibrium DF for a halo having an embedded
disk.  The inversion formula requires both the halo density and the
total central attraction of the halo and disk (assumed spherical).
One must check that the resulting isotropic DF is positive over the
entire energy range, as inversion (also described by \citealt{BT08})
does not guarantee that it is.

\item
A third possible procedure, which we prefer here, is to use halo
compression. One starts with a simple halo model having a known DF,
and computes a revised DF after the disk is added, assuming the
potential change caused by adding the disk was adiabatic.  The method
was pioneered by \citet{Youn80}, who used the fact that the DF,
expressed as a function of the actions, is invariant during an
adiabatic change to the potential.  The two actions in a perfectly
spherical model are angular momentum and radial action.  \citet{SM05}
describe the calculation details for the case of a halo+disk.  As
for Eddington inversion, the attraction of the disk must be assumed
spherical, but \citet{SM05} showed this was an excellent
approximation.  Even if the DF before compression was isotropic,
inserting a disk and/or bulge gives the compressed DF a mild radial
bias.

\end{enumerate}

In addition to being adaptable, halo compression is guaranteed to
yield a physically acceptable model in which the density remains
positive everywhere.  Moreover, the iteration quickly converges to the
equilibrium model, and therefore the before-compression model can be
tweaked if the post-compression model is not to one's liking.

\subsection{Halo + disk models}
Here we adopt the cored isothermal sphere for the pre-compression halo
which has the radial density profile
\begin{equation}
  \rho(r) = {V_0^2 \over 4\pi Gr_c^2} {3 + x^2 \over (1 + x^2)^2},
\end{equation}
where, $x=r/r_c$, with $r_c$ being the core radius, and $V_0$ the
asymptotic circular speed when $x \gg 1$.  An isotropic DF for this
mass profile is readily determined by Eddington inversion.

We truncate this infinite mass halo by limiting the maximum apocentric
distance of any orbit to be $r_t$.  Thus the maximum allowed energy of
an orbit of total angular momentum $L$ is $E_{\rm max} = \Phi(r_t) +
(L/r_t)^2/2$, with $\Phi(r) = \left(V_0^2/ 2\right) \ln( 1 + x^2 )$
being the gravitational potential of the halo before the addition of
a disk.

We insert the exponential disk (eq.~\ref{eq.expdisk}) whose center
coincides with the center of the halo, and compute the resulting rotation
curve of the combined equilibrium model.  Some examples are shown in
Fig.~\ref{fig.RCs}.

\begin{table}
\caption{Numerical parameters for our 3D simulations.}
\label{tab.3Dpars}
\begin{tabular}{@{}lrr}
                   & Cylindrical grid       & Spherical grid \\
\hline
Grid size          & $(N_R,N_\phi,N_z)\quad$ \\
                   &  $ = (170,256,125)$    & $n_r = 201$ \\
Angular compnts    & $0\leq m \leq 8$       & $0 \leq l \leq 4$ \\
Outer radius       & $6.30R_d$              & $45R_d$ \\
$z$-spacing        & $0.025R_d$ \\
Softening rule     & cubic spline           & none \\
Softening length   & $\epsilon = 0.05R_d$  \\
Number of particles & $6\times 10^6$        & $5\times 10^6$ \\

Longest time step  & $0.1(R_d^3/GM)^{1/2}$ \\
Time step zones    & 5 \\
\hline
\end{tabular}
\end{table}

\begin{figure}
\begin{center}
  \includegraphics[width=\hsize,angle=0]{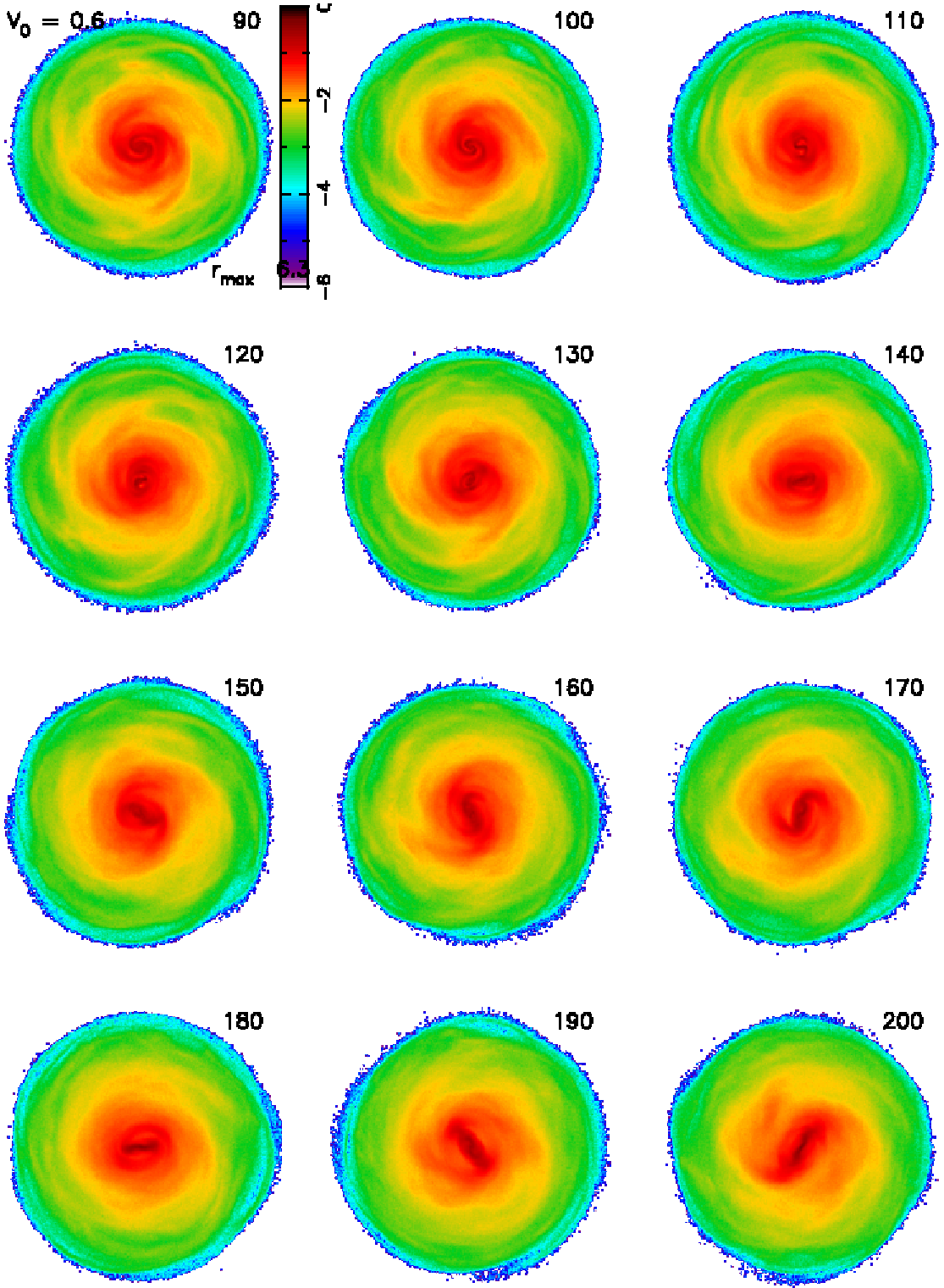}
% jasmine:/data/sellwood/5700/5835/picts.s 5835
\end{center}
\caption{The later part of the evolution of the disk component in the
  $V_0=0.6$ 3D simulation, whose rotation curve is given in the left
  panel of Fig.~\ref{fig.RCs}.  Notice that a strong bar forms.}
\label{fig.run5835}
\end{figure}

\subsection{Evolution of a live halo model}
We select the model shown in the left panel of Fig.~\ref{fig.RCs},
which has a total rotation curve that approximates that of our model
A, and evolve it using our hybrid grid method.  The disk particles are
assigned to a 3D polar grid, while the field of the halo particles is
computed using a multipole expansion on the spherical grid, as is
fully described in \citet{Sell14}.  Our chosen values for the
numerical parameters are given in Table~\ref{tab.3Dpars}.

The evolution of the disk component of this model having a live halo,
which forms a strong bar by $t\sim 150$ is presented in
Fig.~\ref{fig.run5835}.  This contrasts with the model shown in
Fig.~\ref{fig.run5772} which had a very similar (but not identical)
rigid halo.  As has been found previously, the live halo encourages
the formation of the bar.

\begin{figure}
\begin{center}
  \includegraphics[width=\hsize,angle=0]{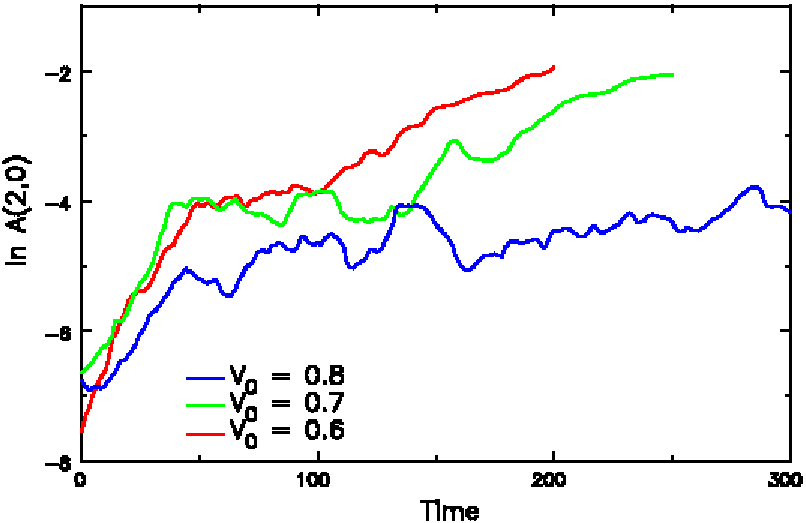}
% jasmine:/data/sellwood/5700/5835/ampllive.s 
\end{center}
\caption{The amplitude evolution of the bar-like logarithmic
  spiral component in the disks of three simulations having
  different compressed live halos.}
\label{fig.ampllive}
\end{figure}

Fig.~\ref{fig.ampllive} gives the amplitude evolution of the bar-like
logarithmic spiral component (eq~\ref{eq.logspi}) of this and two
additional simulations having heavier halos, also shown in
Fig.~\ref{fig.RCs}. That with $V_0=0.7$ for the uncompressed halo also
formed a bar, while only multi-arm spirals and no bar formed in the
model having the still heavier halo with $V_0=0.8$.

\begin{figure}
\begin{center}
  \includegraphics[width=\hsize,angle=0]{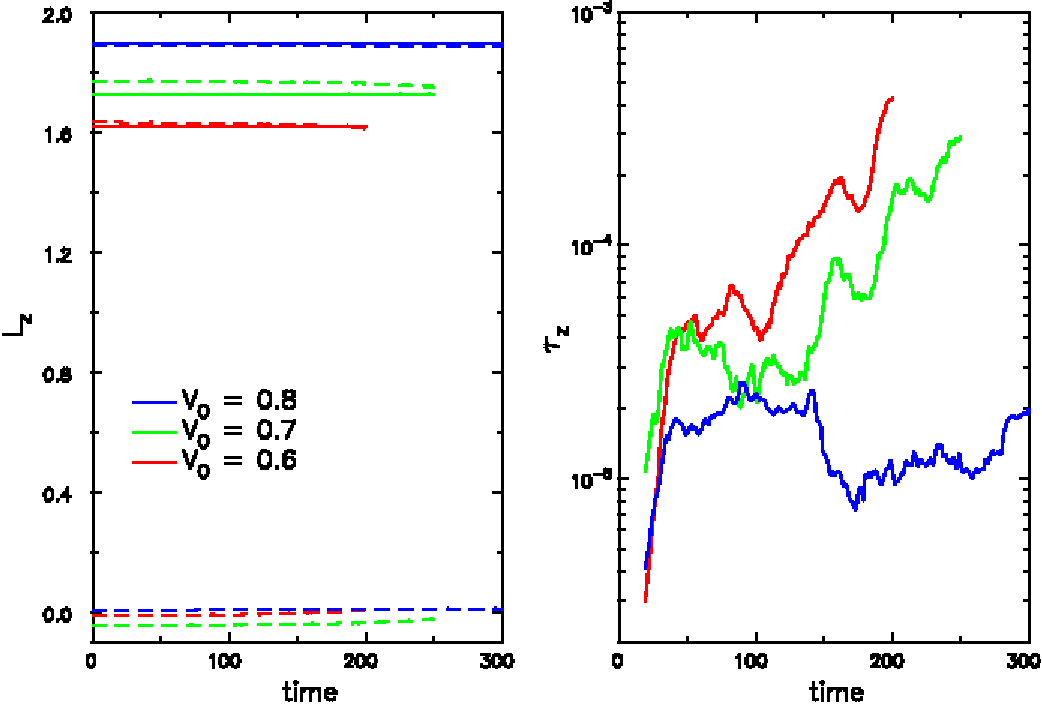}
% jasmine:/data/sellwood/5700/5835/torque.s 
\end{center}
\caption{The dashed lines in the left panel indicate the separate
  angular momenta of the disks and halos, the solid lines are their
  sum, demonstrating that this global integral is well conserved.
  The right panel gives the rates at which angular momentum is taken
  from the disk component and added to the halo.}
\label{fig.torque}
\end{figure}

The left panel of Fig.~\ref{fig.torque} reports the angular momenta of
the separate disk and halo components in these three simulations,
while the right panel gives the torque acting on the halo in all three
cases. The angular momentum taken from the disk is the time integral
of the torque, which is very small in the case of the densest halo
(blue lines).  The formation of a bar is associated with angular
momentum transfer to the halo, as has been reported before
\citep{Sell15}.

In the light of our experience reported in \S2 above, we zeroed out
all bisymmetric force terms on both grids and reran the simulation
illustrated in Fig.~\ref{fig.run5835}, to search for possible more
slowly growing instabilities, but found none.  Thus the question of
possible mode interference in this live halo model does not arise.

\subsection{Disk stability}
The black curves in Fig.~\ref{fig.RCs} indicate $V_{c, \rm max} \simeq
1.10$ for the bar-stable model in the right hand panel and $V_{c, \rm max}
\simeq 1.03$ for the bar unstable model in the middle
panel.\footnote{Note that $V_{c, \rm max}$ includes both the
compressed halo and disk, while here $V_0$ is for the uncompressed halo
only.}  These two results, which are from just two simulations, are
consistent with the old stability criterion proposed by ELN, which
proposed that models having $V_{c, \rm max} \ga 1.1$ should be stable.
Our finding that the ELN stability criterion, which was derived from
2D simulations in rigid halos, holds in these two cases of 3D discs in
live halos was a surprise since \citet{Atha08} argued that it needed
to be revised.

\citet{Toom81} found that the swing-amplifier, which drives the bar
mode in disks, would die away when his parameter $X \ga 3$.
Unfortunately, $X \equiv 2\pi R/m\lambda_{\rm crit}$ is a
locally-defined parameter, but we find that it has a minimum near
$R=R_d$ in our models, where $X_{\rm min} \simeq 2.60$, $3.02$, and
$3.45$ when $V_0 = 0.6$, 0.7 and 0.8, respectively.  The $V_0 = 0.7$
model, for which $X_{\rm min} \simeq 3.02$, was unstable, but it is on
the boundary given by Toomre for effectiveness of the swing amplifier,
and the mild help from halo coupling probably tipped it into instability.
The outcomes in the two other cases were also consistent with the
predictions of swing-amplifier theory.

%%%%%%%%%%%%%%%%%%%% Section 4 %%%%%%%%%%%%%%%%%%
\section{Conclusions} \label{sec.concl}
We have shown that moderately heavy disks embedded in rigid halos can
appear to be stable when disturbance forces within the disk plane are
unrestricted.  This result contrasts both with the findings of
\citet{SC23} and violates the stability criterion proposed long ago by
\citet{ELN}.  We have demonstrated that non-linear scattering by other
modes halts the growth of the bar mode prior to its expected
saturation in these models.  We were unaware at the time of our
earlier study (SC23) that the responsible modes, which have different
rotational symmetries, have linear growth rates that are comparable
to, or even exceed, that of the bar mode.  Since linear perturbation
theory neglects changes to the background disk, the linear growth of a
mode is disturbed when the background disk is altered by the
saturation of a faster-growing mode, which in our models prevented the
incipient bar instability from creating a bar.

The situation is different when disturbance forces are restricted to a
single sectoral harmonic, since all modes having other angular
symmetries are suppressed, and the dominant mode of the selected
rotational symmetry can win out, as was true for all the simulations
in SC23.

While the above discussion applies to disks embedded in rigid halos,
the stability properties of disks in live halos can differ again
because the bar mode elicits a supporting response from the halo even
in the linear regime \citep{Sell15}.  We have also shown that a disk
that appeared to be stable when embedded in a rigid halo was strongly
unstable in a similar halo that was composed of mobile particles.
Furthermore, the stability criteria proposed for disks in rigid halos
by ELN and by \citet{Toom81} appear to hold for our three live halo
models.

Our discovery that bar-forming modes can be inhibited by faster
growing spiral modes has turned out in this study to be an interesting
side issue that ultimately did not affect global stability when
simulated with unrestricted disk forces and embedded in live halos.
If live halo models having more vigorous spiral modes exist, they may
provide an interesting new solution to the long-standing puzzle
presented by the bar instability.

\section*{Acknowledgements}
We thank the referee for a supportive review, thoughtful comments, and
for encouraging us to clarify some technical points.  JAS acknowledges
the continuing hospitality and support of Steward Observatory.

\section*{Data availability}
The data from the simulations reported here can be made available on
request.  The simulation code and analysis software can be downloaded
in one bundle from {\tt http://www.physics.rutgers.edu/galaxy}, and is
documented in the code manual \citep{Sell14}.

\bibliography{54years.bib}{}

\end{document}